\documentclass[conference]{IEEEtran}

\usepackage{cite}
\usepackage{amsmath,amssymb,amsfonts}
\usepackage{algorithmic}
\usepackage{graphicx}
\usepackage{textcomp}
\usepackage{xcolor}
\usepackage{enumitem}
\usepackage{balance}
\usepackage{url}

\begin{document}

\title{Approximate Multiplier Induced Error Propagation in Deep Neural Networks}

\author{
\IEEEauthorblockN{A. M. H. H. Alahakoon}
\IEEEauthorblockA{
\textit{The University of Sydney}\\
Sydney, Australia \\
hala0793@uni.sydney.edu.au}
\and
\IEEEauthorblockN{Hassaan Saadat}
\IEEEauthorblockA{
\textit{University of New South Wales}\\
Sydney, Australia \\
h.saadat@unsw.edu.au}
\and
\IEEEauthorblockN{Darshana Jayasinghe}
\IEEEauthorblockA{
\textit{The University of Sydney}\\
Sydney, Australia \\
darshana.jayasinghe@sydney.edu.au}
\and
\IEEEauthorblockN{Sri Parameswaran}
\IEEEauthorblockA{
\textit{The University of Sydney}\\
Sydney, Australia \\
sri.parameswaran@sydney.edu.au}
}

\maketitle

\begin{abstract}

Deep Neural Networks (DNNs) rely heavily on dense arithmetic operations, motivating the use of Approximate Multipliers (AxMs) to reduce energy consumption in hardware accelerators. However, a rigorous mathematical characterization of how AxMs error distributions influence DNN accuracy remains underdeveloped.
This work presents an analytical framework that connects the statistical error moments of an AxMs to the induced distortion in General Matrix Multiplication (GEMM). Using the Frobenius norm of the resulting error matrix, we derive a closed form expression for practical DNN dimensions, that demonstrates the distortion is predominantly governed by the multiplier’s mean error (bias).
To evaluate this model in realistic settings, we incorporate controlled error injection into GEMM and convolution layers and examine its effect on ImageNet-scale networks. The predicted distortion correlates strongly with the observed accuracy degradation, and an error-configurable AxM case study implemented on an FPGA further confirms the analytical trends. By providing a lightweight alternative to behavioral or hardware-level simulations, this framework enables rapid estimation of AxM impact on DNN inference quality.

\end{abstract}

\maketitle



\section{Introduction}

 The widespread adoption of deep neural networks (DNNs) has driven an exponential rise in computational and energy demands, making efficient arithmetic units essential \cite{computerpower}. Approximate multipliers have emerged as a promising alternative to accurate, hardware-intensive multipliers in DNN accelerators, offering substantial gains in area, delay, and power \cite{Approx2}. Notably, several approximate designs achieve DNN prediction accuracy comparable to that of exact multipliers, despite introducing arithmetic errors \cite{DNNApprox}.

 However, the underlying reason why some approximate multipliers preserve DNN accuracy while others do not has not been formally understood. Existing studies largely rely on empirical evaluation \cite{Approxtrain, TFapprox, REALM}, providing no principled basis for predicting the impact of accuracy on multiplier error characteristics. In this paper, we introduce a mathematical analysis that links approximate multiplier error properties directly to their resulting matrix-level distortion in General Matrix Multiplication (GEMM) operations, a key factor influencing DNN accuracy. This framework identifies the error metrics most critical to preserving model performance and provides systematic guidance for designing approximate multipliers optimized for DNN workloads.

\begin{figure}[htbp]
  \centering
  \includegraphics[width=0.35\textwidth]{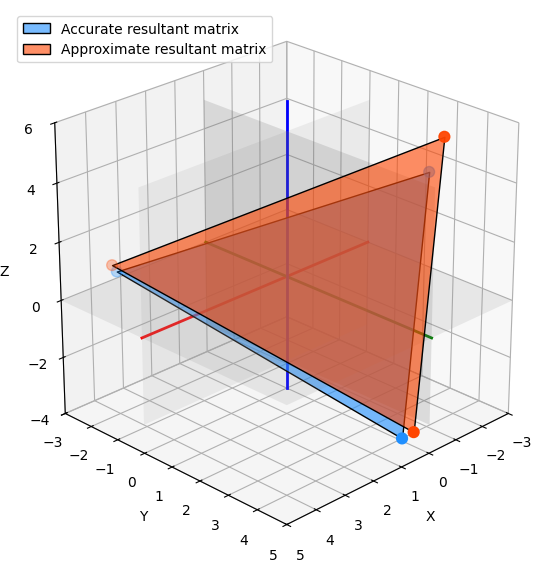}
  \caption{Distortion of the resultant matrix in $\mathbb{R}^{3}$ under numerical error}
  \label{fig:3d}
\end{figure}

 The core computational workload in deep neural networks (DNNs) is dominated by GEMM operations. When approximate multipliers are used within GEMM, the resulting output matrix deviates from the accurate product due to the numerical errors introduced by approximation (Fig.~\ref{fig:3d}). This deviation propagates through the network, distorting the feature space and ultimately causing accuracy degradation. Thus, preserving the shape of the output matrix (maintaining its structural consistency with the accurately computed version) is essential for achieving near-zero accuracy loss when deploying approximate multipliers in DNN computations.

 Motivated by this, our mathematical reasoning introduces a Frobenius norm based metric that quantifies the degree of shape distortion between two matrices: the approximate GEMM output and its accurate counterpart. Then by extending this distortion metric across all GEMM operations in a DNN, we obtain a single accumulated measure that characterizes the network-level impact of AxMs as a whole rather than limiting the analysis to an isolated GEMM.
 This shape distortion metric is derived so that it forms a formal expression that links directly to the error properties of the underlying approximate multiplier. 
 This provides a principled tool for evaluating whether a given approximate multiplier is suitable for DNN workloads, offering clear guidance to designers seeking to ensure accuracy preserving approximation. 
 Our primary contributions are as follows:

 \begin{itemize}[noitemsep,topsep=0pt,leftmargin=*]
 
     \item We present the first formal mathematical framework that analytically links approximate-multiplier error characteristics to their impact on GEMM computations, and extend this formulation across all GEMMs in a DNN to obtain a network-level distortion metric.

     \item We show that, for realistic DNN matrix dimensions, the distortion between approximate and accurate GEMM outputs is dominated by the multiplier’s mean error (bias), while error variance plays only a secondary role.

     
     \item We demonstrate that the proposed theory reliably predicts DNN accuracy trends across multiple ImageNet scale architectures, and validate these predictions on real hardware using an error configurable AxM integrated into Gemmini \cite{GEMMINI} systolic array based DNN accelerator prototyped in an FPGA. 

     \item Building on this analysis, we introduce a lightweight synthetic error injection method that provides fast and architecture-agnostic evaluation of AxMs, providing early insight into their expected DNN accuracy impact without the overhead of full behavioral or hardware-level testing.

     \item Finally, this is the first work that demonstrates that there is \textbf{high fidelity \cite{DBLP:conf/iccad/JavaidIP10} between the calculated Frobenius Norm and the accuracy of networks}, and this fidelity can then be used to predict accuracy without resorting to expensive simulations.
 \end{itemize}

 \section{Background and Related work}

 \subsection{GEMM operation: Building block of DNN computations}

Modern DNNs \cite{lecun2015deep} are equipped with different types of layers spanning convolutional layers to fully connected layers. These fully connected (FC) layers are directly map to matrix multiplications, while convolutional layers that form feature spaces from spatial data are also lowered to GEMM through the widely used im2col transformation, which reshapes overlapping input patches into a matrix and unrolls convolution kernels into another.

This transformation unifies the computations of both FC and convolution layers around the GEMM primitive. 
In addition, the shapes of the matrices involved in these DNN computations are extremely large.  Table~\ref{tab:gemm-sizes} summarizes the matrix sizes that are multiplied in selected layers in popular DNNs.


 \begin{table}[h]
\centering
\caption{Representative GEMM dimensions from major CNN architectures after im2col lowering.}
\begin{tabular}{l|c|c}
\hline
\textbf{Model} & \textbf{Layer} & \textbf{GEMM Size (n × m × p)} \\
\hline
VGG-16\cite{VGG16}   & Conv3\_1  & $25088 \times 576 \times 64$ \\
ResNet-18\cite{resnet}  & Conv1    & $150528 \times 147 \times 64$ \\
ResNet-34  & Conv2\_1 & $50176 \times 576 \times 64$ \\
ResNet-50  & Conv3\_1 & $802816 \times 576 \times 64$ \\
MobileNet-V2\cite{mobilenet} & Conv dw+pw block & $50176 \times 1152 \times 128$ \\
\hline
\end{tabular}
\label{tab:gemm-sizes}
\end{table}


 At such scales, if an AxM is used to compute the GEMM operation, even a comparably small systematic numerical error introduced by the AxM can compound across layers to produce distortions in the DNN’s feature space ultimately degrading accuracy. 
 This risk grows with the inherent properties of DNNs and the employed GEMM computations, making it difficult to predict which approximate multipliers will remain safe for deployment.
 
 Therefore, mathematically grounded error analysis of large-scale matrix multiplication and it's error propagation through DNNs under AxMs is critical when evaluating or designing AxMs for DNN computations. In addition, such mathematical analysis must explicitly connect the statistical error properties of an approximate multiplier to the resulting distortions in GEMM, providing designers with principled guidance rather than heuristic trial and error.

\subsection{Approximate Multipliers for DNN computations}

The use of methods like precision scaling and computation reduction in DNNs demonstrates that DNN computations are particularly an error resilient application \cite{Approx2}. The study \cite{zhao2025insights} demonstrates the potential of having approximation techniques in Large Language models. Therefore, replacing of accurate multipliers with AxMs in DNNs is a promising path for achieving significant energy and area reductions. 



Numerous studies have demonstrated that well-designed approximate multipliers can preserve near-identical accuracy during DNN inference \cite{MBM, 60,49,10}.
To evaluate these multipliers, prior studies typically rely on low level error metrics such as mean absolute error (MAE), mean error (bias error), peak absolute error etc. which provide a coarse characterization of the error distribution introduced by the AxM. 

Despite this empirical success, there remains no rigorous mathematical explanation for why certain AxMs perform remarkably well in DNNs, while others do not \cite{DNNApprox}. In the absence of such understanding, designers are forced to rely on heuristics or expensive, simulation-in-the-loop validation, making design-space exploration slow and inefficient.

Our work provides this missing analytical foundation. We introduce a formal mathematical analysis that links the low-level error characteristics of an individual approximate multiplier to the large-scale distortion introduced in DNN GEMM computations. This framework yields clear, actionable insights by identifying the precise error metrics that must be controlled to maintain high-end-to-end DNN accuracy. To calculate the accuracy of the network, the authors of previous works have relied on time-consuming simulations. In particular, when choosing an approximate multiplier among others, they have had to simulate extensively, often taking hours or even days. This work demonstrates that these multipliers can be quickly selected by calculating the Frobenius Norm of the error matrix.

 \section{Mathematical Error Analysis}

 \subsection{Individual Approximate Multiplier Error Model}

 Let the product of accurate multiplication $z$, between 2 operands $x,y$ be 
 \begin{equation}
     z = x\times y
     \label{1}
 \end{equation}
 and the corresponding approximate multiplication (denoted by $\otimes $) result $c'$ is the sum of the accurate product and an error term, $\varepsilon$:
\begin{equation}
    z'=x\otimes y
    \label{2}
\end{equation}
 \begin{equation}
     z' = z+\varepsilon
     \label{3}
 \end{equation}

The error $\varepsilon$ introduced by an approximate multiplier is a deterministic function of its inputs $(x, y)$. 
Therefore if $x$ and $y$ are samples drawn from a random distribution, the induced error $\varepsilon$ also behaves as a random variable. Hence, we model the error term $\varepsilon$ for each multiplication as a random variable drawn from an error distribution $\mathcal{D}$. 


 This distribution $\mathcal{D}$ is characterized by its first two statistical moments:
 \begin{equation}
      \mu=\mathbb{E}\left[ \varepsilon \right]
      \label{eq:mean}
 \end{equation}
\begin{equation}
     \sigma^{2} = \mathbb{E}[(\varepsilon-\mu)^{2}]
     \label{eq:var}
\end{equation}
In equations \ref{eq:mean} and \ref{eq:var}, if $N$ is the number of sampled error instances $\varepsilon_{i}$, then $\mu$ corresponds to the mean or the bias error of the approximate multiplier $\frac{1}{N}\sum_{i=0}^{N}\varepsilon_{i}$, and $\sigma^{2}$ corresponds to the variance or the spread of error $\frac{1}{N}\sum_{i=0}^{N}(\varepsilon_{i}-\mu)^{2}$ in the error distribution $\mathcal{D}$.

\subsection{GEMM error formulation}

Let matrix $A\in \mathbb{R}^{n\times m}$ and $B\in \mathbb{R}^{m\times p}$. Then the accurate matrix product of $A$ and $B$ is $C=AB$ $(C\in \mathbb{R}^{n\times p})$. Each element in $C$ can be written as,

\begin{equation}
    c_{ij} = \sum_{k=1}^{m} a_{ik} \times b_{kj}
    \label{6}
\end{equation}
Correspondingly, if the matrix product from approximate multiplication is $C'$, $C'\in \mathbb{R}^{n\times p}$ each element in $C'$ can be written as,
\begin{equation}
    c'_{ij} = \sum_{k=1}^{m} a_{ik}\otimes  b_{kj}
    \label{7}
\end{equation}
then by equations \ref{1},\ref{2} and \ref{3},
\begin{equation}
    c'_{ij} = \sum_{k=1}^{m}((a_{ik} \times b_{kj}) + \varepsilon_{ikj})
    \label{8}
\end{equation}
The $\varepsilon_{ikj}$ is the specific error instance for the $ k$th approximate multiplication in the approximate matrix multiplication.

Then the error matrix $E$  $(E\in \mathbb{R}^{n\times p})$ is defined as the element-wise difference between the approximate and accurate resultant matrices.
\begin{equation}
    E=C'-C
    \label{9}
\end{equation}
If each element of the error matrix $E$ is $e$, then from equations \ref{6}, \ref{8} and \ref{9}, 
\begin{equation}
    e_{ij} = \sum_{k=1}^{m}\varepsilon_{ikj}
    \label{10}
\end{equation}

\subsection{Frobenius Norm as a Metric for Matrix Shape Deviation}

The Frobenius norm of a matrix $G\in \mathbb{R}^{n\times p}$ can be introduced as,
\begin{equation}
    \left\| G \right\|_{F}= \sqrt{\sum_{i=1}^{n}\sum_{j=1}^{p}g_{ij}^{2}}
    \label{11}
\end{equation}
Frobenius norm corresponds to the square root of the sum of squared matrix entries and is widely used in linear algebra to measure the magnitude of a given matrix. Therefore if there are 2 matrices in the vector space with same exact elements but has different shapes for example, $A=\begin{pmatrix}
1 & 0 \\
0 & 1
\end{pmatrix}$ and $B=\begin{pmatrix}
0 & 1 \\
1 & 0
\end{pmatrix}$ it has the same Frobenius norm or the same magnitude. 

However, in the formulation of the error matrix $E$ in Equation \ref{9}, when doing the GEMM operation, the positions of the elements that are multiplied are fixed in the vector space (Equations \ref{6} and \ref{7}). Therefore in this case, the Frobenius norm of E captures the global distortion of the approximate multiplied resultant matrix $C'$ with accurate multiplied resultant matrix $C$ and correlates with changes in the geometric  orientation of the output feature space in a DNN computation under AxMs. Taking into account Equations \ref{9}, \ref{10} and \ref{11} it can be derived that the squared Frobenius norm of the error matrix $E$ as,
\begin{equation}
    \left\| E \right\|_{F}^{2}= \sum_{i=1}^{n}\sum_{j=1}^{p}\left( \sum_{k=1}^{m} \varepsilon_{ikj}^{}\right)^{2}
    \label{12}
\end{equation}

Since $\varepsilon$ is from a error distribution $\mathcal{D}$ with mean $\mu$
and variance $\sigma^2$ (Equations \ref{eq:mean},\ref{eq:var}) to further develop the analytical model the expected value of the squared Frobenius norm of error matrix, ($\mathbb{E} \left[ \left\| E \right\|_{F}^{2} \right]$) should be calculated. By the linearity of expectation, 
\begin{equation}
    \mathbb{E} \left[ \left\| E \right\|_{F}^{2} \right] = \sum_{i=1}^{n}\sum_{j=1}^{p} \mathbb{E}\left[  \left( \sum_{k=1}^{m} \varepsilon_{ikj}^{}\right)^{2}\right]
    \label{13}
\end{equation}

and by the assumption of the error terms are pairwise uncorrelated we obtain,
\begin{equation}
    \mathbb{E}\left[  \sum_{k=1}^{m} \varepsilon_{ikj} \right] = \sum_{k=1}^{m}\mathbb{E}\left[ \varepsilon_{ikj} \right]=m\mu
    \label{14}
\end{equation}
\begin{equation}
    Var  \left( \sum_{k=1}^{m} \varepsilon_{ikj} \right)  = \sum_{k=1}^{m}Var\left(\varepsilon_{ikj} \right)=m\sigma^{2}
    \label{15}
\end{equation}

In terms of DNNs, this uncorrelated assumption is a simplification as correlated inputs like convolution kernels can induce correlation in the error stream. However, this first-order model is sufficient to reveal the dominant error scaling properties for GEMM with approximate multipliers.

Then by the alternative variance formula, ($Var(X) = \mathbb{E}\left[ X^{2} \right]-(\mathbb{E}[X])^{2}$) and form Equations \ref{14} and \ref{15},

\begin{equation}
    \mathbb{E}\left[  \left( \sum_{k=1}^{m} \varepsilon_{ikj}^{}\right)^{2}\right] = m\sigma^{2}+m^2\mu^{2}
    \label{16}
\end{equation}
Therefore form Equations \ref{13} and \ref{16} we obtain,
\begin{equation}
    \mathbb{E} \left[ \left\| E \right\|_{F}^{2} \right] = \sum_{i=1}^{n}\sum_{j=1}^{p}\left( m\sigma^{2}+m^2\mu^{2} \right)
    \label{17}
\end{equation}

\begin{equation}
    \mathbb{E} \left[ \left\| E \right\|_{F}^{2} \right] = np\left( m\sigma^{2}+m^2\mu^{2} \right)
    \label{18}
\end{equation}

As discussed earlier, when using approximate multipliers in GEMM operations, preserving the shape of the resultant matrix is crucial. The Frobenius norm of the error matrix $E = C' - C$ provides a measure of how much the approximate result $C'$ deviates from the accurate matrix $C$ in terms of overall geometric structure. Therefore, minimizing $\mathbb{E}\left[ \|E\|_{F}^{2} \right]$ in Equation \ref{18} directly corresponds to preserving the feature-space geometry required for correct DNN inference.

From Equation \ref{18}, $\mathbb{E}\left[ \|E\|_{F}^{2} \right]$ decomposes into the variance component ($npm\sigma^{2}$) and the bias component ($npm^2\mu^2$).
Both terms scale linearly with outer dimensions $n$ and $p$, but differ in how they scale with the inner dimension $m$. The variance term grows as $O(m)$, whereas the bias term grows as $O(m^2)$. Therefore, whenever
\begin{equation}
     \left| \mu \right|\gg \frac{\sigma}{\sqrt{m}} \Rightarrow  (npm^{2}\mu^{2} \gg npm\sigma^2)  
    \label{19}
\end{equation} 
the bias term dominates the total distortion.

For realistic DNNs, this is usually the case because $m$ is a very large value (Table~\ref{tab:gemm-sizes}) making $\frac{\sigma}{\sqrt{m}}$  very small. This implies that even a small non-zero bias $\mu$ in the multiplier error distribution can dominate the total matrix distortion in large-scale matrix multiplications.

\subsection{Accumulated Expected Frobenius Norm for DNNs}

The Equation~\ref{18} demonstrates the distortion for just a single GEMM operation, but a DNN is a sequence of such GEMMs from convolutions converted through im2col and fully connected layers. Therefore, under the same abstraction of having errors from a distribution of mean $\mu$ and variance $\sigma^2$ the network level accumulated expected Frobenius norm can be demonstrated by the sum of layer contributions.

Let $\mathbb{E} \left[ \left\| E \right\|_{F}^{2} \right]_{l}$ be the Expected squared Frobenius norm for a layer $l$ with GEMM dimensions of $n_{l}, p_{l}$ and $m_{l}$ then by Equation ~\ref{18},

\begin{equation}
    \mathbb{E} \left[ \left\| E \right\|_{F}^{2} \right]_{l} = n_{l}p_{l}\left( m_{l}\sigma^{2}+m^2_{l}\mu^{2} \right)
    \label{20}
\end{equation}

Therefore the accumulated expected squared Frobenius norm for a DNN with $T$ GEMM operation layers  boils down to,

\begin{equation}
    \mathbb{E} \left[ \left\| E \right\|_{F}^{2} \right]_{net} = \sum_{l=1}^{T}\left(  n_{l}p_{l}\left( m_{l}\sigma^{2}+m^2_{l}\mu^{2} \right)\right )
    \label{21}
\end{equation}

However, the Eq.~\ref{21} captures the distortion before applying activation functions.
But common activation blocks such as ReLU are element-wise 1-Lipschitz $\left| f(a)-f(b) \right|\le \left| a-b \right|$ and the Frobenius norm is the $L2$ norm over all elements. Therefore,

\begin{equation}
    \left\| f(X+E) -f(X) \right\|_{F}\le \left\| E \right\|_{F}
    \label{22}
\end{equation}

Hence, Eq.~\ref{21} provides a conservative upper bound on the post nonlinearity matrix distortion due to numerical errors introduced by AxMs.

\begin{figure*}[h]
    \centering
    \setlength{\tabcolsep}{1pt}
    \begin{tabular}{ccc}
        \parbox[c]{0.32\linewidth}{\centering \textbf{ResNet18 Fig 1a}\\\fbox{\includegraphics[width=\linewidth, height=5.5cm]{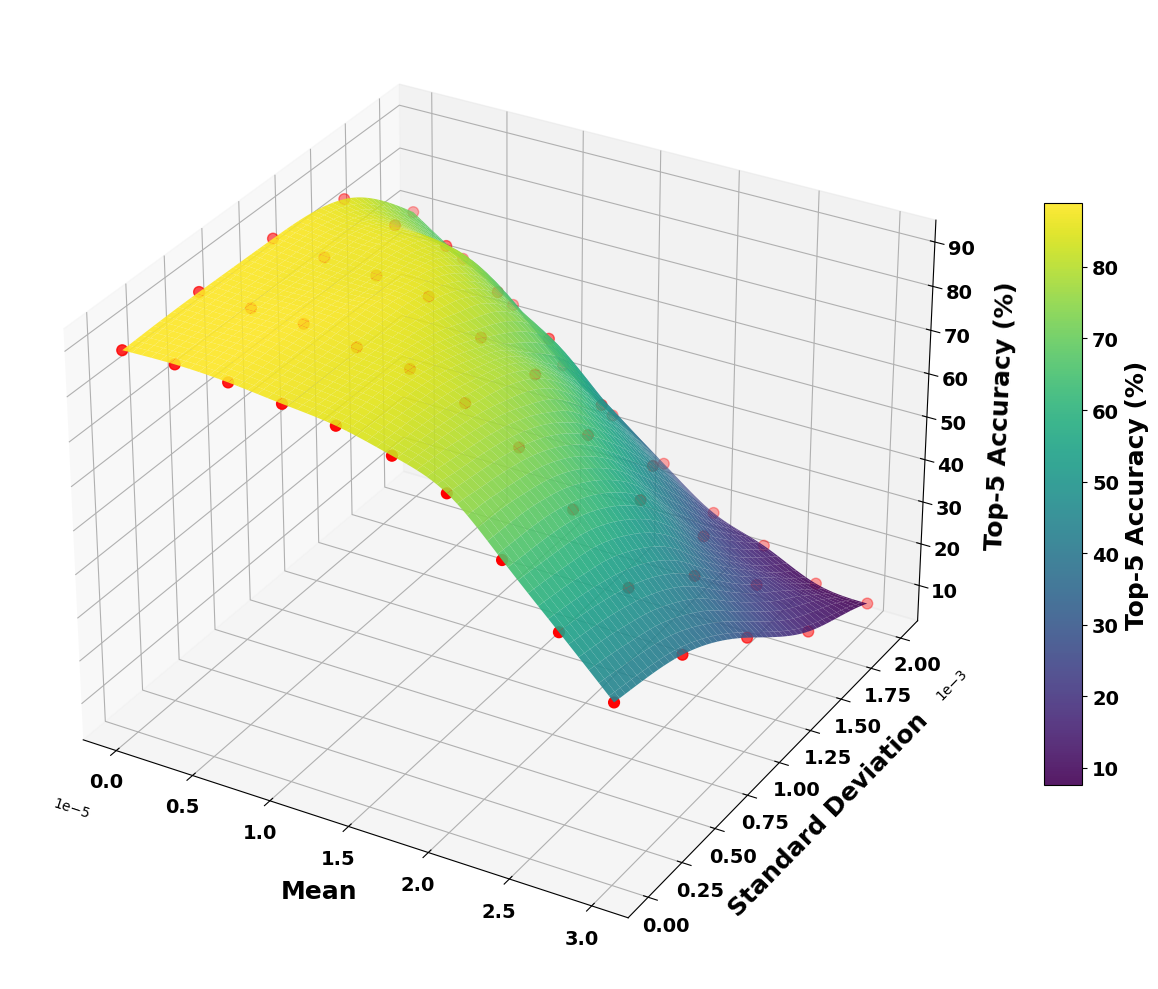}}} &
        \parbox[c]{0.32\linewidth}{\centering \textbf{ResNet18 Fig 1b}\\\fbox{\includegraphics[width=\linewidth, height=5.5cm]{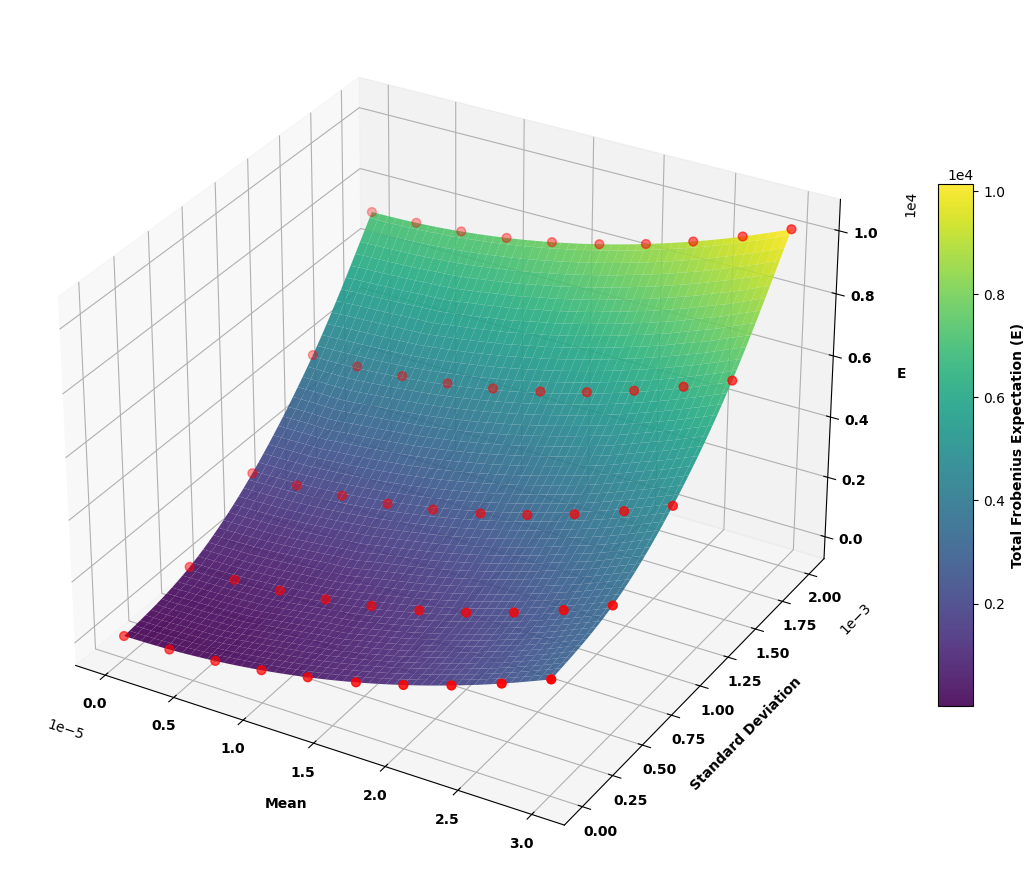}}} &
        \parbox[c]{0.32\linewidth}{\centering \textbf{ResNet18 Fig 1c}\\\fbox{\includegraphics[width=\linewidth, height=5.5cm]{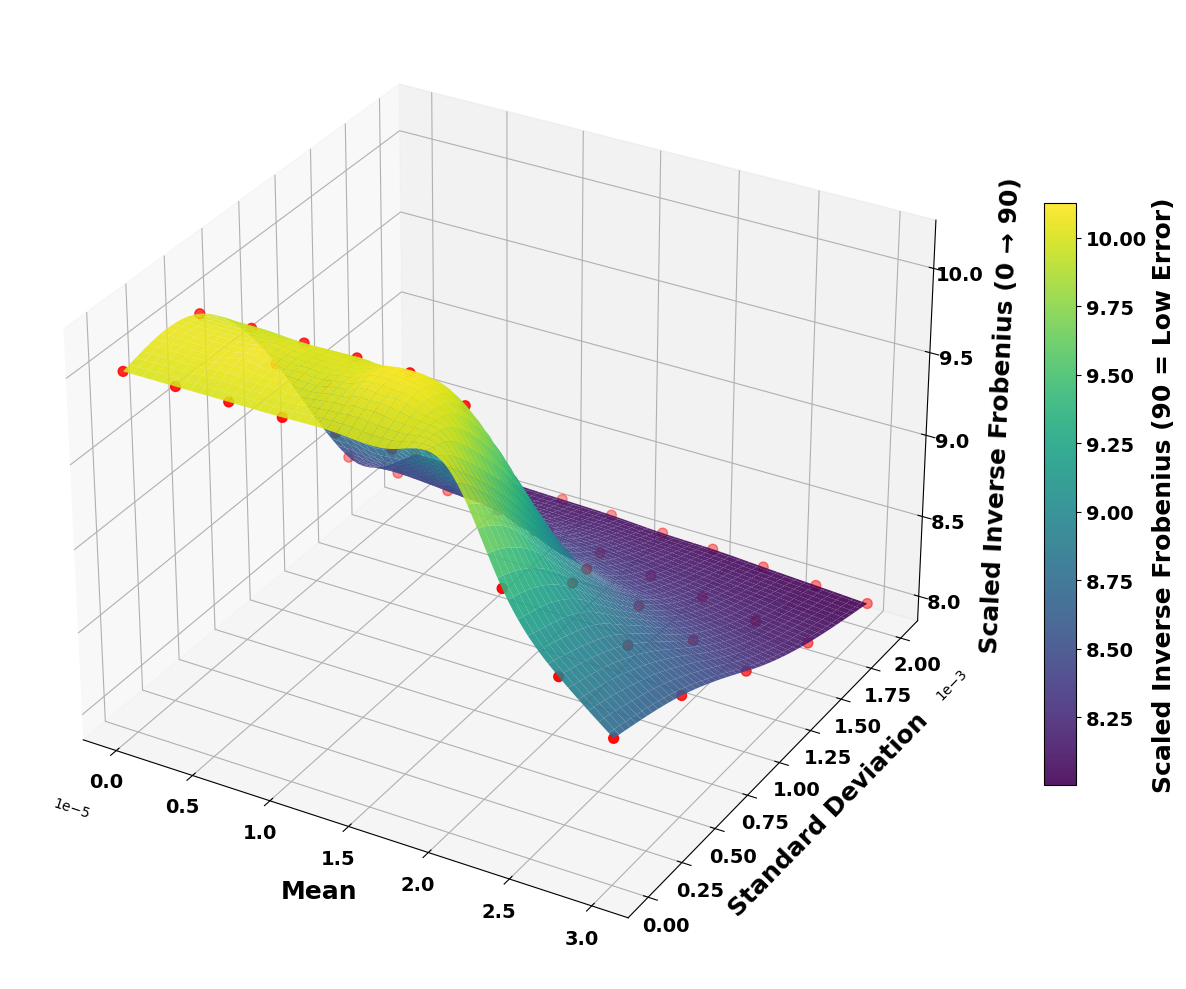}}} \\
        
        \\[1mm]

        \parbox[c]{0.32\linewidth}{\centering \textbf{ResNet34 Fig 2a}\\\fbox{\includegraphics[width=\linewidth, height=5.5cm]{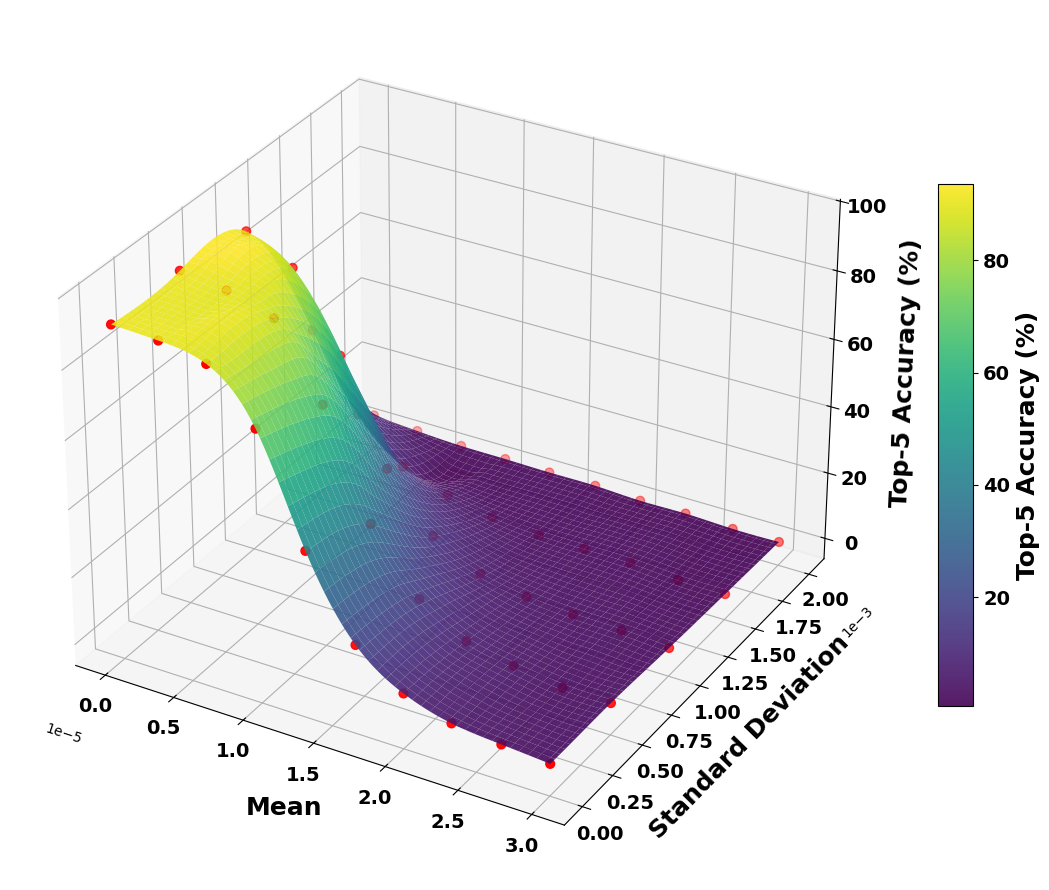}}} &
        \parbox[c]{0.32\linewidth}{\centering \textbf{ResNet34 Fig 2b}\\\fbox{\includegraphics[width=\linewidth, height=5.5cm]{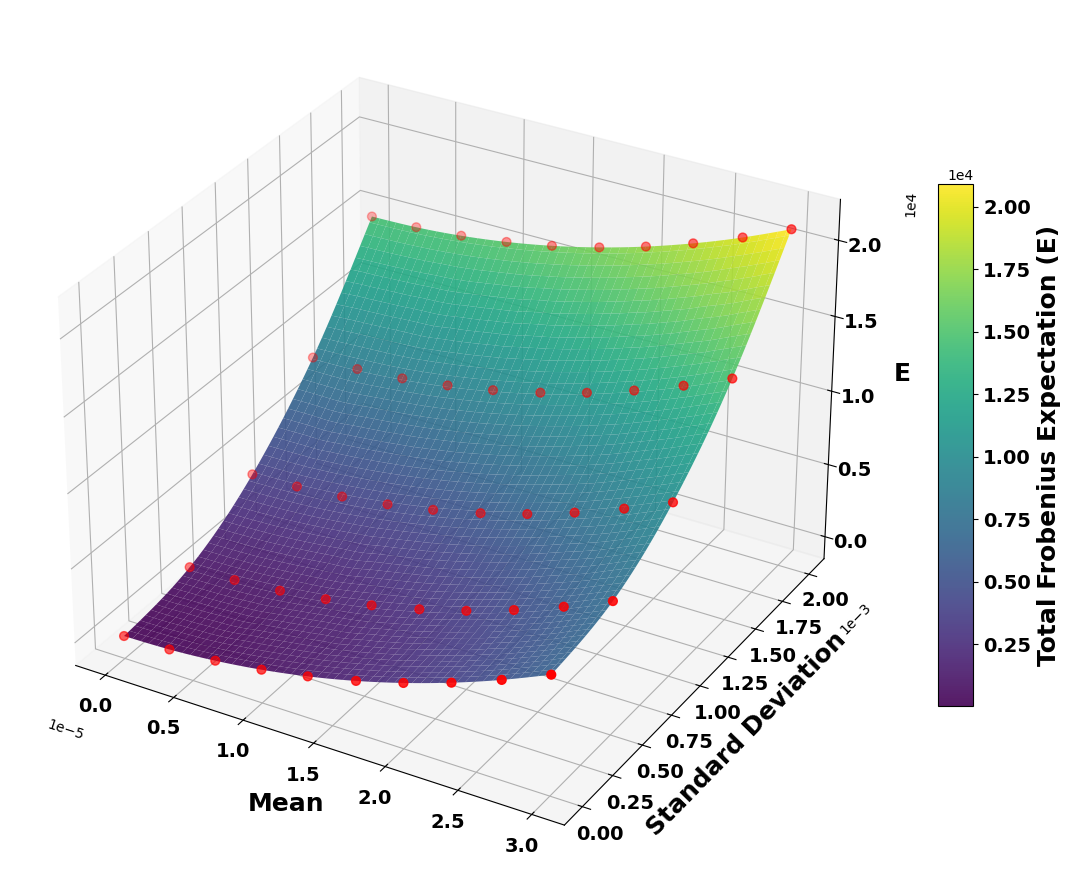}}} &
        \parbox[c]{0.32\linewidth}{\centering \textbf{ResNet34 Fig 2c}\\\fbox{\includegraphics[width=\linewidth, height=5.5cm]{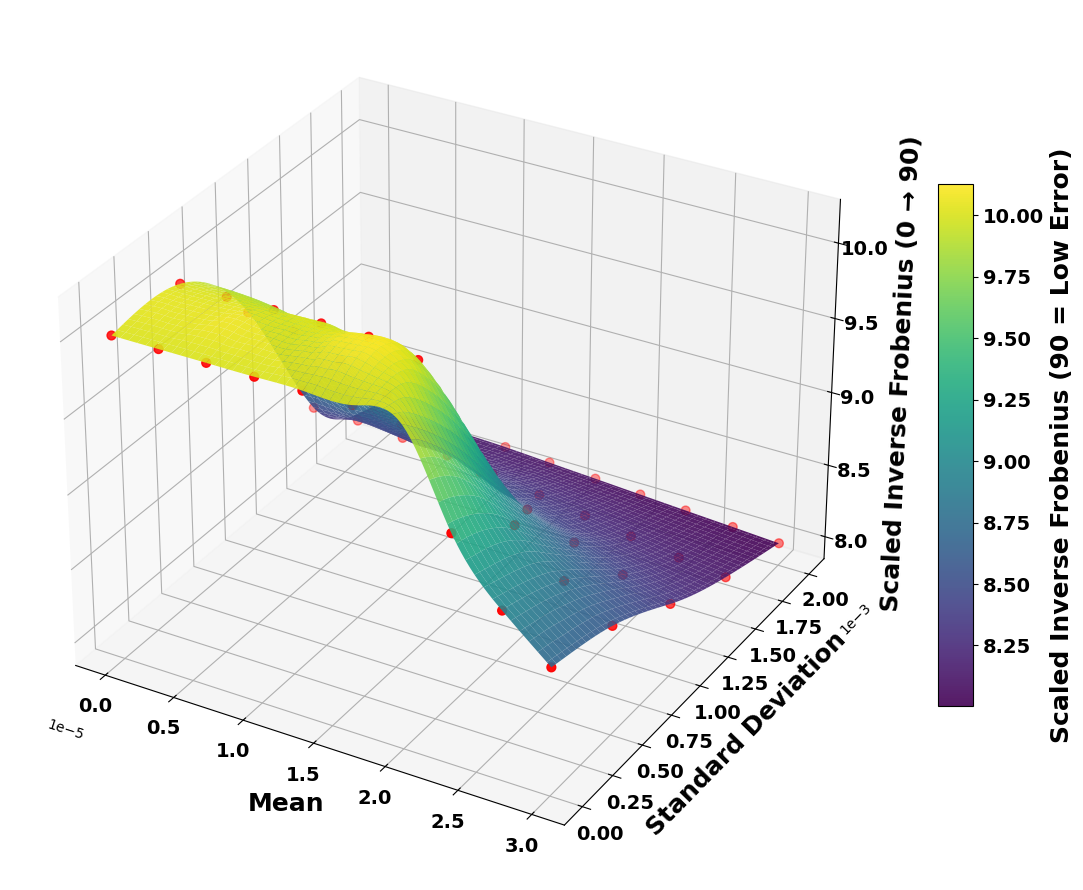}}} \\

        \\[1mm]

        \parbox[c]{0.32\linewidth}{\centering \textbf{ResNet50 Fig 3a}\\\fbox{\includegraphics[width=\linewidth, height=5.5cm]{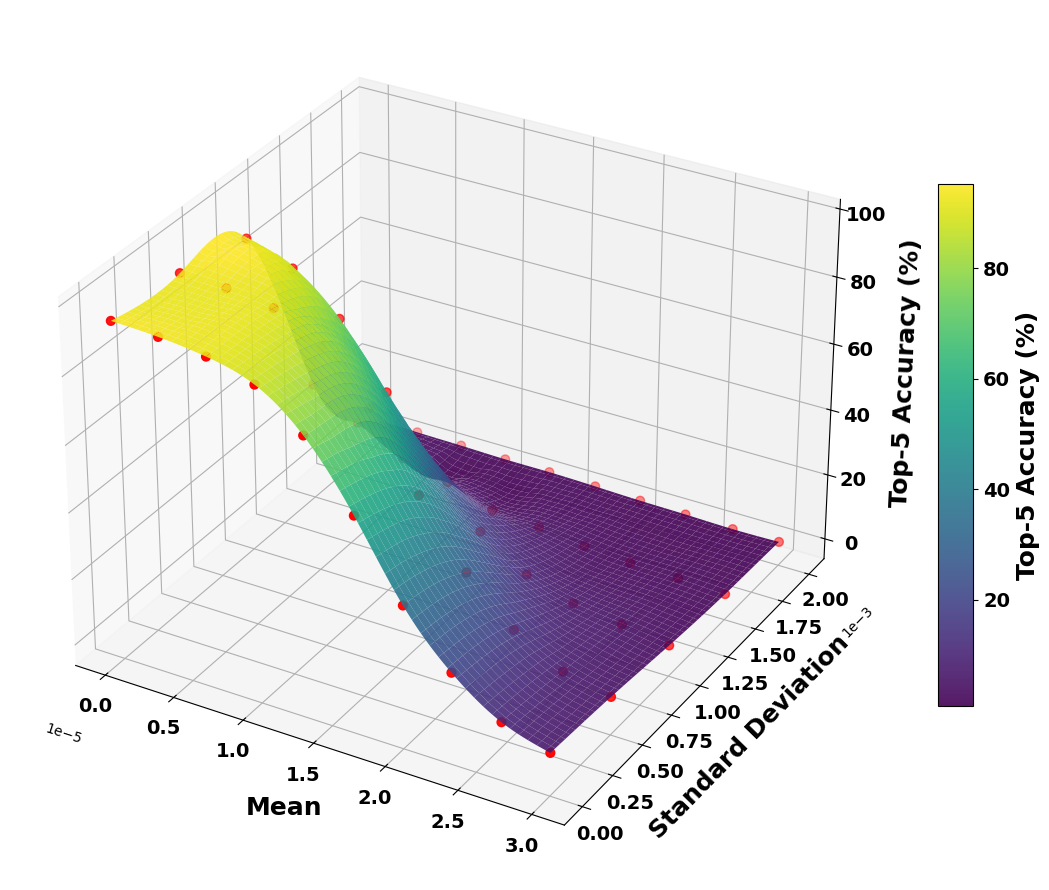}}} &
        \parbox[c]{0.32\linewidth}{\centering \textbf{ResNet50 Fig 3b}\\\fbox{\includegraphics[width=\linewidth, height=5.5cm]{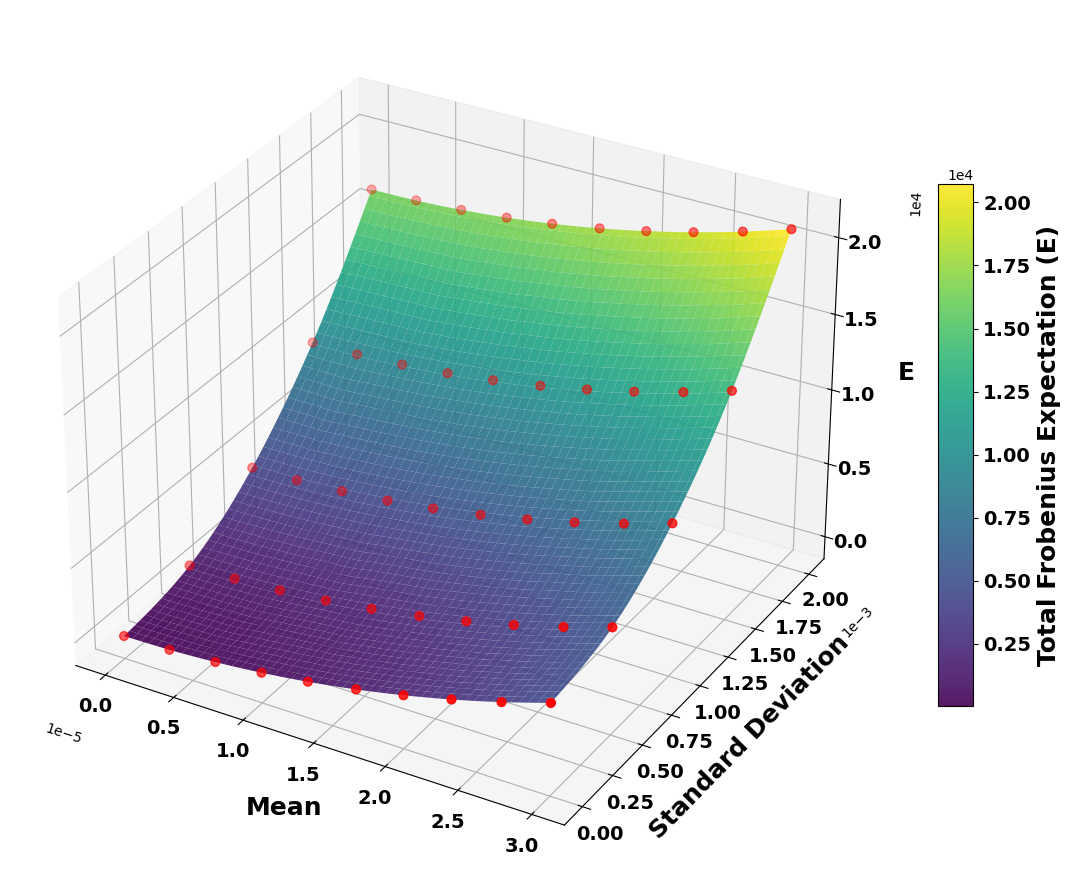}}} &
        \parbox[c]{0.32\linewidth}{\centering \textbf{ResNet50 Fig 3c}\\\fbox{\includegraphics[width=\linewidth, height=5.5cm]{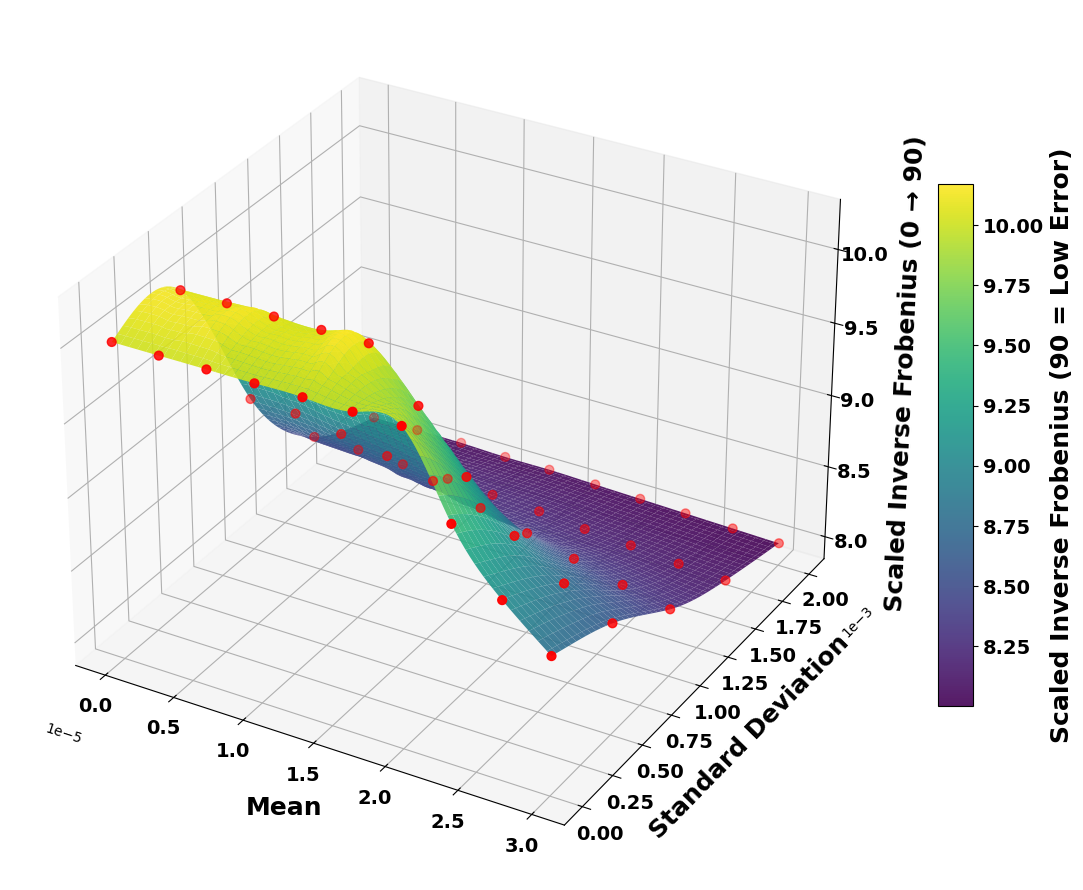}}}
    \end{tabular}
    \caption{Model based variation of accuracy with synthetic numerical error injection. Column a: Top-5 Accuracy. Column b: Accumulated Expected Squared Frobenius Norm. Column c: Inverse Frobenius Norm (Scaled and Capped).}
    \label{fig:9grid}
\end{figure*}

\section{Experimental Setup}

For the evaluation and validation of our mathematical error analysis experiments were structured on 2 main components. First, a software-based synthetic error-injection validation setup was employed and then a hardware-based validation using an AxM implemented on a DNN accelerator and prototyped on a an FPGA was employed. Together, these two components offer both exhaustive analytical validation and practical hardware relevance.

\subsection{Software based synthetic error injection setup}

The mathematical analysis we developed, models the effects of using approximate multipliers on GEMM operation.
To demonstrate that these findings and their assumptions are valid within complete DNNs, which introduce non-linearities (e.g., ReLU) and other complex layers, we designed an experimental controlled error-injection framework.
This framework introduces a numerical error into each multiplication between the 2 individual elements getting multiplied by GEMM operation (Equation \ref{8}). This error is sampled from a normal distribution $N(\mu, \sigma)$ where the user can control $\mu$ and $\sigma$ to achieve different multiplier error behaviors.
This independent control of $\mu$ and $\sigma$ is critical for verifying our mathematical model but is impossible to achieve with real approximate multiplier hardware, as any given physical design has fixed error properties. Therefore, having this synthetic error injection model allows us to directly test the soundness of our mathematical error model and, most importantly, validate our hypothesis that the matrix distortion is bias-dominated.

This error injection was tested out for trained weights of ResNet-18, ResNet-34 and ResNet-50 \cite{resnet} on the Imagenet \cite{imagenet} validation dataset. The error parameters were independently varied over the ranges of $\mu \in \left[ 0, 3\times 10^{-5} \right], \sigma\in \left[ 0,2\times 10^{-3} \right]$. These ranges were selected to reflect the error characteristics observed by performing Monte Carlo analysis in real approximate multiplier designs reported in prior hardware literature \cite{MBM,REALM,60}. All experiments were executed on an NVIDIA RTX 3090 GPU.

\subsection{Hardware based validation setup}

In order to have hardware based validation, we use the Minimally Biased Multiplier (MBM) \cite{MBM}, a design that approximates only the mantissa multiplication by adding an error correction to Mitchell's multiplier \cite{mitchell2009computer} (a log based approximation) while preserving accurate exponent handling. 
Unlike fixed behavior approximate multipliers, the MBM is implemented in an error configurable form, allowing its error correction term to be deliberately varied while keeping the underlying structure unchanged. 
This characteristic makes the MBM an ideal approximate multiplier for validating our mathematical analysis in a controlled environment. 
We have successfully integrated MBM into the Gemmini systolic array \cite{GEMMINI} and deployed the design on a Digilent Genesys2 FPGA board and executed MobileNet~\cite{mobilenet} inference on the ImageNet dataset.
In order to get the characterization parameters of the MBM’s error distribution, mean and standard deviation Monte Carlo simulations were performed using $2^{24}$ uniformly distributed random inputs chosen to match the typical dynamic range of DNN weights and activations, allowing the resulting mean and standard deviation to reflect realistic inference conditions..

The MBM's internal error correction term was changed from 0 (no error correction) to it's native design setting (1010). This gave a range of $(\mu, \sigma)$ values from the Monte Carlo characterization.


\section{Results and Discussion}

\subsection{Error injection and validation}

 In the selected mean and standard deviation parameter space for the synthetic error injection setup, we observed clear and consistent accuracy degradation, with top-5 accuracy dropping from the high 80\% range to below 10\% depending on the chosen $\mu$ and $\sigma$. 

The Fig.~\ref{fig:9grid} summarizes the effect of controlled multiplier error on DNN inference. The first column of subplots shows the variation of the top-5 accuracy of ResNet-18/34/50 as the independent variables mean and standard deviation is varied.
The second subplot shows the corresponding accumulated expected squared Frobenius norm (Equation~\ref{21}) computed from converting the architectural parameters of network’s convolutional layers and FC layers into GEMM operations.
The third subplot shows a scaled version of the inverse of accumulated expected squared Frobenius norm (capped at a threshold value of 10 for visual clarity) against the mean and variance of the error distribution.

Across these three visualizations it can be observed that accuracy collapses precisely in the regions where the accumulated expected squared Frobenius norm becomes large, validating our analytical model. When $\mathbb{E} \left[ \left\| E \right\|_{F}^{2} \right]_{net}$ is minimal, all three ResNet variants maintain high accuracy, but as distortion increases either through higher variance or more aggressively through higher mean the accuracy drops sharply from the high 80\% range to below 10\%. This behavior directly reflects the exact information from Equation~\ref{18}. Moreover, deeper networks such as ResNet-50 degrade earlier because their larger bias scaling factor (Equation~\ref{18}) amplifies the perturbation more severely than in ResNet-18 or ResNet-34. Overall, this figure demonstrates that the accumulated Frobenius distortion is an effective, architecture-aware predictor of DNN robustness under approximate multipliers, and that bias dominated error is the principal driver of catastrophic accuracy loss when using approximate multipliers.

To further quantify the strength of the relationship between the accumulated Frobenius distortion ($\mathbb{E} \left[ \left\| E \right\|_{F}^{2} \right]_{net}$) and DNN accuracy, we computed the Spearman rank correlation between them across all tested mean and standard deviations.
The Spearman correlation captures monotonic relationships without assuming linearity, and is therefore robust to the nonlinear relationship observed between distortion due to numerical errors and accuracy. 
The resulting spearman correlation coefficient was –0.862 that indicates a strong negative monotonic correlation (–1 denotes perfect inverse correlation and +1 denotes perfect direct correlation). Therefore as accumulated squared Frobenius distortion increases, accuracy consistently decreases. This high-magnitude correlation provides an additional statistical confirmation that our analytical metric is tightly aligned with real DNN behavior.

\subsection{Hardware Evaluation Using an Error configurable approximate multiplier}

To complement the synthetic error-injection results, we now evaluate the MBM an error configurable approximate multiplier, when deployed inside a real DNN accelerator prototyped on an FPGA. 

For each configuration of the MBM’s error correction parameter, and their mean and variance of the different error distributions the accumulated expected squared Frobenius distortion predicted by our Equation~\ref{21}.


\begin{table}[h]
\centering
\caption{Hardware evaluation of MBM on Gemmini: measured $\mu$ and $\sigma$, predicted Expected squared Frobenius distortion, and ResNet-50 accuracy}
\label{tab:mbm_fpga}
\begin{tabular}{c|c|c|c|c}
\hline
\textbf{Bias} & $\boldsymbol{\mu}$ ($\times 10^{-5}$) & $\boldsymbol{\sigma}$ ($\times 10^{-3}$) & 
$\boldsymbol{\mathbb{E}[||E||_F^2]}$ ($\times 10^{4}$) & \textbf{Acc. (\%)} \\
\hline
0000 & -1.850 & 26.200 & 280.732 & 72.4 \\
0001 & -1.730 & 25.200 & 259.702 & 72.9 \\
0010 & -1.240 & 23.200 & 220.064 & 73.8 \\
0011 & -1.120 & 22.400 & 205.140 & 74.1 \\
0100 & -0.627 & 20.600 & 173.462 & 74.8 \\
0101 & -0.510 & 19.900 & 161.868 & 75.0 \\
0110 & -0.021 & 18.400 & 138.374 & 75.8 \\
0111 & 0.097 & 17.900 & 130.956 & 75.6 \\
1000 & 0.604 & 16.800 & 115.374 & 76.3 \\
1001 & 0.721 & 16.600 & 112.652 & 76.8 \\
1011 & 1.310 & 16.200 & 107.350 & 77.2 \\
\textbf{1010} & 1.200 & 16.100 & 106.016 & \textbf{77.5} \\
\hline
\end{tabular}
\end{table}

Table~\ref{tab:mbm_fpga} summarizes the measured error statistics, predicted distortion values, and corresponding hardware accuracies. The Table ~\ref{tab:mbm_fpga} clearly demonstrates that increasing hardware bias leads to a proportional growth in the predicted Frobenius distortion, which in turn aligns tightly with the observed accuracy drop on the FPGA. 
Moreover, the Spearman’s rank correlation coefficient between the predicted distortion and measured accuracy is close to -1, indicating a near-perfect inverse monotonic relationship.
This strong alignment between theory and hardware based measurements reinforces our central claim that the Frobenius-based error metric is a reliable predictor of real-world accuracy degradation across different approximate multiplier configurations.

\subsection{Runtime Comparison of Synthetic Error injection, GPU simulations and FPGA emulations}

In order to assess the practical evaluation cost of our synthetic error injection analysis, we compared the per-batch inference latency of ResNet-50 on ImageNet using commonly used evaluation methods of behavioral simulation of approximate multipliers through custom GPU-based CUDA kernels \cite{Approxtrain,TFapprox} and FPGA emulation on the Gemmini DNN accelerator using the approximate multipliers. 

The behavioral simulation \cite{Approxtrain} required 301ms per batch, while the full FPGA-based evaluation incurred a substantially higher latency of 4567ms per batch. In contrast, our synthetic error-injection method completed the same workload in only 31ms (since these were done on native multipliers as opposed the one in \cite{Approxtrain} which had to have specialized software multipliers to simulate the approximate multipliers) per batch. 

These results demonstrate that our mathematical error analysis is not only theoretically sound but also directly applicable to rapidly assessing approximate multiplier designs in practical scenarios. By accurately reflecting the accuracy trends observed in both behavioral and FPGA evaluations, synthetic error injection enables fast, reliable, and scalable design-space exploration for AXMs.

Additionally, the Frobenius Norm of the error has high fidelity to the network's accuracy (Spearman's correlation greater than -0.85). By simply calculating the Frobenius norm we can make predictions about the accuracy without spending enormous amounts of time simulating the system.

%



\section{Conclusion and Future work}
This work introduced a formal mathematical analysis that characterizes how numerical errors generated from approximate multipliers propagate through GEMM and accumulate across an entire DNN. Our analysis reveals that, for realistic DNN dimensions, distortion is fundamentally bias-dominated, providing a clear design principle for developing accuracy-preserving approximate multipliers. The theory was validated through both large-scale synthetic error-injection experiments and real hardware evaluation using an error-configurable AxM integrated into the Gemmini DNN accelerator prototyped on an FPGA.

Looking forward, we aim to build a complete design and evaluation framework for approximate multipliers in DNN accelerators, using the proposed metric as its analytical foundation. This includes automating AxM characterization, integrating fast synthetic accuracy prediction, and supporting rapid hardware–software co-exploration. In addition, the mathematical approach presented here can be extended beyond approximate multiplication: future work will investigate analogous formulations for other approximation techniques such as \textbf{pruning, quantization, mixed-precision arithmetic, and compressed GEMM kernels}. Together, these directions move toward a unified theoretical foundation for understanding and designing approximation mechanisms in modern DNN accelerators.


\newpage
\balance


\bibliographystyle{ieeetr}
\bibliography{References}

\end{document}